\documentclass[conference]{IEEEtran}
\usepackage{amssymb}  % assumes amsmath package installed
\usepackage{amsmath,enumerate} % assumes amsmath package installed
\usepackage{graphicx} % for pdf, bitmapped graphics files

\newcommand{\msu}{\mathcal{U}}
\newcommand{\msv}{\mathcal{V}}

\newcommand{\hspm}{\hspace{-2pt}}

\newcommand{\acl}{\bar{A}}
\newcommand{\numn}{\boldsymbol{n}}
\newcommand{\numm}{\boldsymbol{m}}

\newcommand{\numnone}{\boldsymbol{n-1}}
\newcommand{\numr}{\boldsymbol{r}}
\newcommand{\om}{\omega}
\newcommand{\kont}{\mathit{Con}}

\newcommand{\numro}{\boldsymbol{r_0}}
\newcommand{\numno}{\boldsymbol{n_0}}

\begin{document}
\title{On Pole Placement and Invariant Subspaces}
%:\\ Pole Placement Revisited}

\author{\IEEEauthorblockN{Naim Bajcinca}
\IEEEauthorblockA{Max Planck Institute for \\  Dynamics of Complex Technical Systems\\
Sandtorstr. 1, 39106, Magdeburg, Germany}
\tt\small Email:  bajcinca@mpi-magdeburg.mpg.de%}
\vspace{-12pt}
}

\maketitle

%%%%%%%%%%%%%%%%%%%%%%%%%%%%%%%%%%%%%%%%%%%%%%%%%%%%%%%%%%%%%%%%%%%%%%%%%%%%%%%%
\begin{abstract}
The classical eigenvalue assignment problem is revisited in this
note. We derive an analytic expression for pole placement
which represents a slight generalization of the celebrated Bass-Gura and Ackermann formulae, and also is closely related to the modal procedure of Simon and Mitter.
\end{abstract}
%%%%%%%%%%%%%%%%%%%%%%%%%%%%%%%%%%%%%%%%%%%%%%%%%%%%%%%%%%%%%%%%%%%%%%%%%%%%%%%%
\IEEEpeerreviewmaketitle

\vspace{-5pt}
%===============================================================================
\section{Introduction}
\label{Introduction}
\vspace{-2pt}
%--------------------------------------------------------------
%The problem of spectrum (or pole) placement has been extensively
%investigated starting with 60' and 70' of the last century

For a single-input linear time-invariant system
$\dot{x}\,=\,Ax\,+\,bu$ with $x\in\mathbb{R}^n$,
$A\in\mathbb{R}^{n\times n}$,  $b\in\mathbb{R}^{n\times 1}$, the
solution to the pole placement problem provides the feedback gain $k\in\mathbb{R}^n$ in $u=k^T x$,
such that the open-loop eigenvalues $\varLambda(A)$ are shifted to
some prespecified values $\varLambda(\bar A)$ where $\bar A :=A+bk^T$,
\cite{Kailat80, Ackermann:1972vl, SimMit68}, etc. In this note, we
utilize a left eigenvector assignment procedure for a
controllable pair $(A,b)$ to derive the following pole placement analytic expression:
\begin{align}
\label{polplac}
k^T = \omega_{n-r}^T q_r(A),%\quad r\in\numno
\end{align}
where $1\leq r\leq n$, $\omega_{n-r}\in\mathbb{R}^n$ and
$q_r(A)\in\mathbb{R}^{n\times n}$ represent the design
parameters that independently assign $(n-r)$ and $r$ eigenvalues,
respectively, and are defined as follows: Let
$q_n(\lambda)=q_{n-r}(\lambda) q_r(\lambda)$ be the specified closed-loop
characteristic polynomial, where $q_{n-r}(\lambda)$ and
$q_{r}(\lambda)$ are real polynomials in $\lambda$ with leading
coefficients equal to one, and let them host the desired $(n-r)$ and $r$
eigenvalues, respectively. Then, $\om^T_{n-r}=\gamma^T_{n-r}T^{-1}$,
where $q_{n-r}(\lambda)=[1,\lambda,\ldots,\lambda^{n-1}]\gamma_{n-r}$,
$\gamma_{n-r}\in\mathbb{R}^{n}$,
and $T$ represents the controller canonical state-space
transformation matrix \cite{Kailat80}, while $q_r(A)$ is the matrix polynomial corresponding to $q_r(\lambda)$.
To the best author's knowledge, \eqref{polplac} has
not appeared in that form previously in the control literature and
could be of interest in the sense that it includes both the Ackermann
and Bass-Gura formulae as special cases. Indeed, it will be shown
later in the paper that for $r=n$ we obtain
the Ackermann formula,
and for $r=0$ we can link \eqref{polplac} to the Bass-Gura formula. We also
stress its close relationship to the procedure of Simon and
Mitter.

%===============================================================================
%\subsection*{Preliminaries}
%\label{Preliminaries}
%--------------------------------------------------------------
\emph{Preliminaries \& Notation:~} $\mathbb{C}_{-}$ stands for the open
left-hand complex half-plane. %The superscript $H$ refers to the
                              %Hermitian transpose, which includes
                              %transposition and complex conjugation.
By $\varLambda(A)$ we denote the
multiset of the eigenvalues of the matrix $A$. $(\lambda,\omega)$ is an eigenpair of $A$
(i.e. $\omega^HA=\lambda\omega^H$) if and only if $(\lambda,Q^{-1}\omega)$ is an
eigenpair of the similar matrix of $Q^{-1}A Q$. A real matrix $A$ can be
factorized into a product $Q^T T Q$, where $Q$ is an orthogonal matrix
and $T$ is lower quasitriangular (i.e., block lower triangular with
$1\times 1$ or/and $2\times 2$ blocks along the diagonal), representing the real
Schur decomposition \cite{qref:matrix-analysis-HJ-vol1}.
$\msv\subseteq\mathbb{C}^n$ is said to be $A-$invariant if
there exists a matrix $Y$ such that $AV=VY$, where
$\msv=\textit{Range}(V)$.  The controllability
matrix $[b,Ab,\ldots, A^{n-1}b]$ of the pair $(A,b)$ is denoted by $\kont(A,b)$.
%$\text{``adj''}$ represents the adjoint operator,
%i.e. $\text{adj}(A)=\det(A) A^{-1}$.
Finally, we use the shorthand:
$\numr:=\{1,\ldots,r\}$ for $r\in\mathbb{N}$ in $i\in\numr$ to indicate
$i\in\{1,\ldots,r\}$; $i\in\numro$ allows $i$ to take also
the value $0$.

\vspace{-2pt}
%=================================================================
\section{Spectrum assignment}
\label{sec:Eigenvalue assignment}
%\vspace{-1pt}
%------------------------------------------------------------------
%\subsection{Derivation}
%\label{Derivation}

Consider the state space representation of a finite-dimensional controllable  single-input linear time invariant  system:
$\dot{x}= Ax+bu$. It is well-known that for any arbitrary multiset  of
{self-conjugate} eigenvalues  $\{\lambda_i\}_{i\in\numn}$ in
$\mathbb{C}_-$, there exists always a unique state feedback gain $k
\in \mathbb{R}^n$ %with $u=k^Tx$
which solves the pole assignment problem \cite{Kailat80}. In the sequel, we provide an original method for computation of $k$.

Let ${\om_{n-1}}\in \mathbb{C}^n$ be a left  eigenvector of the closed-loop system matrix $\acl:=A+b k^T$ corresponding to an arbitrary eigenvalue $\lambda_1 \in \mathbb{C}_-$. Then, with $\om_{n-1}^H(A+ bk^{T})=\lambda_{1}\om_{n-1}^H$, we claim:
\begin{equation}
k^T = {\om_{n-1}^H (\lambda_1 I -A)}\quad\text{and}\quad \om_{n-1}^Hb=1,
\label{eq2:statefeedback}
\end{equation}
whereby in light of implementation, care has to be taken in selecting a pair $\om_{n-1}$ and $\lambda_1$ that guarantee a real outcome $k\in\mathbb{R}^n$.
%Moreover, while the right hand-side in \eqref{eq2:statefeedback} involves complex terms, the produst is still real, due to $\om_{1}^Hb=1$.
Observe, that  the right-hand side statement in
\eqref{eq2:statefeedback} results from the fact that $\om_{n-1}
/\om^H_{n-1} b$ is a left eigenvector of $\acl$, as well, and the
condition $\om_{n-1}^H b \neq 0$, which is guaranteed by the
controllability of the pair $(A,b)$. Indeed, if the opposite would
hold true, i.e. if $\om_{n-1}^Hb=0$, we would have:
$\om_{n-1}^H(A+bk^T)=\om_{n-1}^H A=\lambda_1 \om_{n-1}^H$ for all
$k$, indicating that $\lambda_1$ is an eigenvalue of $A$ and $\acl$
simultaneously, i.e. it cannot be shifted by any $k$, which
contradicts the controllability of $(A,b)$.

Furthermore, equation \eqref{eq2:statefeedback} reveals that the
remainder eigenvalues in the multiset $\{\lambda_i\}_{i=2}^n$ are
uniquely specified by the left eigenvector $\omega_{n-1}$. Hence, it
is natural to pose the spectrum assignment in terms of computing
the eigenvector $\om_{n-1}$ such that a prespecified multiset of
self-conjugate (not necessarily distinct) eigenvalues
$\{\lambda_j\}_{j=2}^n$ are assigned to
\begin{align}
\label{eq:Acl}
\acl= \left(I-{b \om_{n-1}^H}\right)A+\lambda_{1}{b\om_{n-1}^H}.
\end{align}

To this end, we start with the characteristic polynomial of the {closed loop} matrix $\acl$, which (with a little of technical effort) is shown to be given by:
\begin{align}
\label{eq:chpeigs}
\det(\lambda I - \acl )  = (\lambda -\lambda_1){\om^H_{n-1}}\textrm{adj}(\lambda I-A^T)b.
\end{align}
Next, consider the controller canonical form $\dot{\xi}= A_c\xi+b_c u$, with $T A_c= A T$,  $T b_c= b$, and %where we follow the convention with $b_c=[0,\ldots,0,1]^T$.
\begin{displaymath}
A_{c}=\left(\begin{array}{cccc}
0 & 1 &  \cdots & 0 \\
\vdots &   &      &  \\
0 & 0 &  \cdots & 1 \\
-a_{n} & -a_{n-1} & \cdots  & -a_{1}
\end{array}\right),
% use packages: array
\, b_{c}=\left(\begin{array}{l}
0 \\
\vdots \\
0\\
1
\end{array}\right).
\end{displaymath}
Here, $T:=\mathcal{C}\mathcal{C}_{c}^{-1}$ \cite{Kailat80} indicates the transformation $x= T\xi$, where, for convenience, we denote by
$\mathcal{C}:=\kont(A,b)$ and $\mathcal{C}_{c}:=\kont(A_c,b_c)$ the
open-loop and closed-loop controllability matrix \cite{Kailat80}, respectively.
The characteristic polynomial of $A$ then reads:
\begin{equation}
p(\lambda) = \det(\lambda I - A) = \lambda^n+a_{1}\lambda^{n-1}+\ldots+ a_n.
\label{eq:plambda}
\end{equation}
Following the discussion related to equation \eqref{eq2:statefeedback}, if we let
\begin{equation} 	
\label{eqgama10}
\gamma^H_{n-1}:=[\gamma_{n-1,n-1},\ldots, \gamma_{n-1,1},~1]
\end{equation}
represent the desired left eigenvector, and $\lambda_1$ the corresponding eigenvalue of the closed loop
$\acl_{c}:=A_c+b_c k_c^T = T^{-1}\acl T$ %(under the feedback $k_c$!)
in the $\xi$-coordinates, then from \eqref{eq:chpeigs} we get
\begin{equation}
%|\lambda I -  \acl_{c}| = (\lambda-\lambda_{1})\Upsilon^T\hspm(\lambda) \, {\gamma_1},
\det(\lambda I -  \acl_{c}) = (\lambda-\lambda_{1}) {\gamma^H_{n-1}} \, \Upsilon\hspm(\lambda),
\label{canocical char pol}
\end{equation}
where we introduce:
%\[
$\Upsilon\hspm(\lambda): =[1~\lambda~\ldots~\lambda^{n-1}]^T= \textrm{adj} (\lambda I-A_c^T)b_c$.
%\]
From \eqref{canocical char pol} it is obvious that the eigenvalues $\{\lambda_j\}_{j=2}^{n}$ of the closed-loop matrix $\acl_{c}$ (that is, of  $\acl$, as well) are independent of the parameters $a_1, \ldots, a_{n}$, and they are entirely determined by the left eigenvector $\gamma_{n-1}$. % and independent of the original open-loop matrix $A_c$.
On the other hand, let \eqref{canocical char pol}  be specified by a desired closed-loop characteristic polynomial of the form:
\begin{align}\label{eq:descp}
q_n(\lambda) = \lambda^{n}+\alpha_1\lambda^{n-1}\ldots+\alpha_{n}.
\end{align}
Equation \eqref{canocical char pol} says that $\gamma_{n-1}$ hosts the parameters of the polynomial $q_{n-1}(\lambda)$, where
$q_{n}(\lambda)=(\lambda-\lambda_1) q_{n-1}(\lambda)$. Explicitly, it
can be checked that $\gamma_{n-1}$, as defined in \eqref{eqgama10}, is given by the recursive algorithm:
%\begin{align}
%w_{c,1} = \alpha_1+ \lambda_1,~
$\gamma_{n-1,i} = \alpha_i+ \gamma_{n-1,i-1}\lambda_1$ for $i\in\numnone$,
%\label{eq:wcs}
%\end{align}
where, in accordance with our adoption in \eqref{eqgama10}: $\gamma_{n-1,0}=1$. %or, alternatively, by Viete formulae in terms of the desired eigenvalues $\lambda_2,\ldots,\lambda_n$. \\
Moreover, with $\acl$ and $\acl_c$ being similar, we have
\begin{align}
\om_{n-1}^H= \gamma_{n-1}^H \mathcal{C}_c \mathcal{C}^{-1},\quad k^T = \om_{n-1}^H (\lambda_1 I -A).
\label{eq:poleplacement}
\end{align}
%yielding a closed formula for the state-feedback gain:
%using \eqref{eq:Acl}, we have %for $\acl_{c}=T^{-1}\acl T$
%\begin{align}
%\label{eq:Tttt}
%\acl_c =\left(I-\frac{b_c(w^T T)}{(w^T T) b_c}\right)A_c+\lambda_{1}\frac{b_c(w^T T)}{(w^TT)b_c}.
%\end{align}
%Comparing this to the parallel expression \eqref{eq:Acl} in $\xi$-coordinates implies: $w^T T= w_c^T$,
%which in conjunction with \eqref{eq:wcs} leads to the required expression for the state-feedback gain

This represents our initial pole assignment formula. Next, we
generalize it and demonstrate its relationship to the Bass-Gura and
Ackermann formulae. First, it is readily verified that
\vspace{-10pt}
\begin{equation}
\label{eq:bg1}
\gamma_{n-1}^H (\lambda_1 I -A_c) = [\alpha_1-a_1,\ldots,\alpha_n-a_n]:= \gamma_n^H,
\end{equation}
%which we refer to as ``displacing'' the eigenvalue $\lambda_1$ from the pencil $\lambda_1 I -A_c$ into the vector $\gamma_0$.
indicating that all the closed-loop eigenvalues in $\{\lambda_j\}_{{j\in\numn}}$ are ``encoded'' in the (real) vector $\gamma_n$, whereas $\gamma_{n-1}$ carries the information about $\{\lambda_j\}_{j=2}^{n}$. Then, the Bass-Gura formula:
\begin{equation}
\label{eq:bassguraformula}
k^T = \gamma^H_n\mathcal{C}_c \mathcal{C}^{-1}
\end{equation}
results immediately, if we rewrite \eqref{eq:poleplacement} as:
%\begin{equation}
%\label{eq:poleplacement_2}
$k^T = \gamma_{n-1}^H (\lambda_1 I -A_c) \mathcal{C}_c \mathcal{C}^{-1}$,
%\end{equation}
with the term $T^{-1}=\mathcal{C}_c \mathcal{C}^{-1}$ shifted right most. %Notice that the latter expression of $\gamma_n^H$ represents an exception in comparison

Equation \eqref{eq:bg1} can be interpreted as ``pulling out'' or ``carrying over'' the eigenvalue $\lambda_1$ from $\gamma_{n}$ via the factor $\lambda_1 I -A_c$, this necessarily introducing $\gamma_{n-1}$. By proceeding in the same way, one can pullout the eigenvalue $\lambda_2$ from $\gamma_{n-1}$ by means of $\lambda_2I-A_c$, $\lambda_3$ from $\gamma_{n-2}$ via $\lambda_3I-A_c$, and so on. Hence, we can introduce
\begin{align}\label{eq:gammar}
\gamma^H_{n-r}:=[\gamma_{n-r,n-1},\ldots, \gamma_{n-r,r},~1,0,\ldots,0], \quad r\in \numn
\end{align}
using: % On the other hand, ``pulling out'' the eigenvalues $\lambda_2,\ldots,\lambda_{n}$ from $\gamma_{1}$ into the corresponding pencil forms in the latter equation, is equivalent to introducing another vector term $\gamma_{r}$ by
\begin{equation}
\label{eq:poleplacement_4_2}
%\gamma_{1}^H (\lambda_1 I -A_c) =: \gamma_{r}^H~ \prod_{i=1}^{r} (\lambda_iI-A_c),
\gamma_{n}^H =\gamma_{n-r}^H~ \prod_{i=1}^{r} (\lambda_iI-A_c),
\end{equation}
where the $(r-1)$ zeros (for $r\geq 2$) result due to the ``absence''
of the eigenvalues $\lambda_2,\ldots,\lambda_r$ in $\gamma_{n-r}$,
while the $n-r$ non-zero terms carry the information about $\lambda_{r+1},\ldots,\lambda_n$.
In this sense, by substituting \eqref{eq:poleplacement_4_2} into  \eqref{eq:bassguraformula}, our spectrum assignment formula \eqref{eq:poleplacement} can be set in the general form:
\begin{equation}
\label{eq:poleplacement_gen}
k^T = \gamma_{n-r}^H \mathcal{C}_c \mathcal{C}^{-1} \prod_{i=1}^{r} (\lambda_iI-A), \quad r\in \numn,
\end{equation}
%Herein, the information about the $n-r$ desired eigenvalues $\lambda_{r+1},\ldots,\lambda_n$ is captured into $\gamma_{r}$, while the remaining  $r$ ones $\lambda_1,\ldots,\lambda_r$ sit in their corresponding pencils.
which can be slightly generalized to
\begin{equation}
\label{eq:poleplacement_gen_34}
k^T = \omega_{n-r}^H q_r(A), \quad r\in \numno,
\end{equation}
with $q_0(A):=I_n$ and otherwise:
\begin{equation}
\omega_{n-r}^H := \gamma_{n-r}^H \mathcal{C}_c \mathcal{C}^{-1},\quad q_r(A) := \prod_{i=1}^{r} (\lambda_i I -A).
\label{canocical char pol_2_234}
\end{equation}
Clearly, equation \eqref{eq:poleplacement_gen_34} represents the generalized form of our initial expression in \eqref{eq:poleplacement}. For $r \geq 1$ the vector $\gamma_{n-r}$ is simply \emph{defined by the coefficients of the polynomial $q_{n-r}(\lambda)$}, where
\begin{equation}
q_n(\lambda) = q_{n-r}(\lambda)\, q_r(\lambda).
\label{qlambda}
\end{equation}
The definition of $\gamma_{n}$ (i.e. reflecting the Bass-Gura formula with $r=0$, c.f. \eqref{eq:bg1}) \emph{represents an exception} to this rule.

Now, consider  the special case with $r=n$ and let $q_n(A)$ denote the real matrix polynomial corresponding to the desired characteristic polynomial $q_n(\lambda)$ from \eqref{eq:descp}.
%Then, the closed-loop polynomial \eqref{canocical char pol} reads:
%\begin{equation}
%q(A) = \prod_{i=1}^{n} (\lambda I -A).
%\label{canocical char pol_2}
%\end{equation}
 Then, using $\gamma^H_{0} = [1,0,\ldots,0]$ from \eqref{eq:gammar}, and:
%\[
$[1,0,\ldots,0]\cdot \mathcal{C}_c = [0,\ldots,0,1]$,
%\]
we obtain  the Ackermann formula directly from \eqref{eq:poleplacement_gen}:
\begin{align}
\label{eq:ack}
k^T = [0,\ldots,0,1]\, \mathcal{C}^{-1} q_n(A).
\end{align}

\subsection{Comments}

%Few remarks are now in order:

\emph{(i)}~Expressions \eqref{eq:poleplacement_gen}
and \eqref{eq:poleplacement_gen_34} provide a direct link of the
Bass-Gura and Ackermann formulae. Moreover, it represents a
generalization thereof: the former one results with $r=0$ (leading to the definition \eqref{eq:bg1} for $\gamma_{n}$), while the latter one   for $r=n$ in  \eqref{eq:poleplacement_gen}. Notice that from \eqref{eq:ack} we immediately obtain
\[
\omega_0^T=[0,\ldots,0,1]\, \mathcal{C}^{-1}.
\]
%The zeros included hereinn indicate that the pole placement ``authority'' is given to the matrix polynomial $q_n(A)$.

%In this sense, with regard to \eqref{eq:poleplacement_2} and \eqref{eq:poleplacement_4_2}, our pole placement formula \eqref{eq:poleplacement} can be given the general form:
%\begin{equation}
%\label{eq:poleplacement_gen}
%k^T = \gamma_{r}^H \mathcal{C}_c \mathcal{C}^{-1} \prod_{i=1}^{r} (\lambda_iI-A).
%\end{equation}
%Herein, the information about the $n-r$ desired eigenvalues $\lambda_{r+1},\ldots,\lambda_n$ is captured into $\gamma_{r}$, while the remaining  $r$ ones $\lambda_1,\ldots,\lambda_r$ sit in their corresponding pencils.
%

\emph{(ii)}~The desired conjugate eigenpairs should be ``encoded'' jointly in \eqref{eq:poleplacement_gen}, either in the real vector $\omega_{n-r}$ or in the real matrix polynomial $q_r(A)$ to benefit from the numerical computation with real numbers. Therefore, without loss of generality we may consider
\begin{equation}
\label{eq:poleplacement_gen_rer}
k^T = \om_{n-r}^T q_r(A), \quad r\in\numno, %\prod_{i=1}^{r} (\lambda_iI-A).
\end{equation}
as the general form of our spectrum assignment formula. %shall utilize the generalized version of \eqref{eq2:statefeedback} given by \eqref{eq:poleplacement_gen}: where $q_r(A)$ represents the matrix polynomial introduced in the item (iii) and $\om_r\in\mathbb{R}^r$ is assumed (see item (ii)). % To this end, we prefer embedding the information on fixed (i.e. unaltered) eigenvalues in $ % only as $\gamma_r$ in \eqref{eq:poleplacement_gen} is a real vector.
In this sense, it is also convenient to use a real $\lambda_1$ in
\eqref{eq2:statefeedback}.

\emph{(iii)}~If $\om_{n-1}$ in \eqref{eq2:statefeedback} is selected to be the left eigenvector of the open-loop matrix $A$ corresponding to a real eigenvalue, say $\mu_{1}$, then from \eqref{eq:Acl} we have $\acl = A+\Delta_1 {b \om_{n-1}^T}$, with $\Delta_1:=\lambda_1-\mu_{1}$ referring to a real shift. The remainder open-loop eigenvalues $\{\mu_i\}_{i=2}^n$ are thereby unaltered, as for any right eigenvector $\nu_{n-i}$ of $A$ corresponding to the eigenvalue $\mu_i$, we have $\acl \nu_{n-i} = A \nu_{n-i}=\mu_i \nu_{n-i}$, $i\in\{2,\ldots,n\}$ (as a consequence of $\om_{n-1}^T\nu_{n-i}=0$). In this case we retain:
\[
k^T={\Delta_1}\om_{n-1}^T,
\]
which represents the well-known result of Simon and Mitter
\cite{SimMit68} (cf.~{pp.\,338}). It is important to observe in this
case the geometric interpretation of the vector term $\om_{n-1}$ in
\eqref{eq2:statefeedback}: it is {orthogonal to the invariant
  subspace} corresponding to the {eigenvalues that remain
  unchanged}. We discuss this more generally in the next section.
%%%

% n\tred{
% \emph{(iv)}~Formula \eqref{eq:poleplacement_gen} suggests a factorization of the desired characteristic polynomial $q(\lambda) = q_{n-r}(\lambda)\,q_r(\lambda)$, where $q_{n-r}(\lambda)$ and $q_r(\lambda)$ are polynomials of order $n-r$ and $r$ in $\lambda$, respectively. The coefficients of the polynomial $q_{n-r}(\lambda)$ specify the vector $\gamma_{r}$ in \eqref{eq:poleplacement_gen}  by a recursive algorithm analogously to \eqref{eq:wcs}, while the right-hand side product in \eqref{eq:poleplacement_gen} is given by the matrix polynomial $q_r(A)$.
% }

\emph{(iv)}~Finally, due to the presence of the factor
$\mathcal{C}^{-1}$, which for large $n$ is typically ill-conditioned,
related well-known numerical robustness problems are inherent in the
expression \eqref{eq:poleplacement_gen}. In the sequel, we discuss the
avoidance of such difficulties. % is discussed in the next two s. %Section\,\ref{Sequential spectrum assignment}.

%, \cite{KuNicDoo85}.

%(vi)~\tred{multiplicity larger then one, simplicity eigenvalues}

% by utilizing the projection approach from \cite{Saad86}.

%
%
%%------------------------------------------------------------
\subsection{Partial spectrum assignment}
\label{Partial pole placement}
%%------------------------------------------------------------
%\subsection{Sequential spectrum assignment}
%\label{Sequential spectrum assignment}
%%----------------------------------------------------------
Next, we consider the usability of the vector $\om_{n-r}\in\mathbb{R}^n$ in the context of the partial spectrum assignment and a sequential spectrum assignment based thereon, which consists in shifting a submultiset of open-loop self-conjugate eigenpairs, say $M_{r}=\{\mu_i\}_{i=1}^{r}$, to some prescribed self-conjugate $L_{r}=\{\lambda_i\}_{i=1}^{r}$, while keeping the remainder $(n-r)$-ones of $M_{n-r}=\{\mu_i\}_{i=r+1}^n$ unaltered ($r\in\numn$). %Therefore we shall utilize \eqref{eq:poleplacement_gen_rer}.
% \begin{equation}
% \label{eq:poleplacement_gen_rer}
% k^T = \om_{r}^T q_r(A), \quad r\in\numn, %\prod_{i=1}^{r} (\lambda_iI-A).
% \end{equation}
% where $q_r(A)$ represents the matrix polynomial introduced in the item (iii) and $\om_r\in\mathbb{R}^r$ is assumed (see item (ii)). % To this end, we prefer embedding the information on fixed (i.e. unaltered) eigenvalues in $\om_r$.

To this end, consider the operator description of $A$:
\vspace{-5pt}
\begin{align}
\label{eq:geom1}
{A} = \begin{bmatrix} X & 0 \\ \ast & Y \end{bmatrix} : \begin{matrix} \mathcal{U} \\ \oplus \\ \mathcal{V} \end{matrix} \rightarrow \begin{matrix} \mathcal{U} \\ \oplus \\ \mathcal{V} \end{matrix},
\end{align}
\vspace{-5pt}
 corresponding to the real Schur decomposition:
\begin{align}
\label{eq:werew}
A\, (U, V)= (U, V) {\left(\hspm \hspm\begin{array}{cc}
X & 0 \\
\ast &  Y
\end{array}\hspm\hspm \right)},
\end{align}
where $\msu\oplus\msv=\mathbb{R}^n$ (i.e. $\msu$ and $\msv$ are complementary subspaces), $\msu=\textit{Range}(U)\subseteq\mathbb{R}^{r}$, $\msv=\textit{Range}(V)\subseteq\mathbb{R}^{n-r}$ is the $A$-invariant subspace (i.e. $AV=VY$) corresponding to the eigenvalues in $M_{n-r}$, and $[U,V]\in\mathbb{R}^{n \times n}$ is orthogonal (i.e., $\msu$ and $\msv$ are mutually orthogonal subspaces).
Next, introducing %the parametrization
\begin{align}
\label{eq:kappa}
\om_{n-r} = U \eta
\end{align}
in terms of $\eta\in\mathbb{R}^{r}$ in \eqref{eq:poleplacement_gen_rer}, it can be readily checked that the block-triangular form is preserved under feedback \cite{Saad86}:
\begin{align}
\left(\hspm\hspm\begin{array}{c}U^T \\ V^T\end{array}\hspm\hspm\right)
\acl\,\, (U, V) = \left(\hspm\begin{array}{cc}
X\hspm + U^Tb\eta^T q_{r}(X) & ~~0 \\
\ast & ~~Y
\end{array}\right).
\end{align}
Note that due to the re-appearance of $Y$ in the diagonal, the eigenvalues in  $M_{n-r}$ remain unaltered in $\acl$, while those from  $M_{r}$ change subject to the parameter $\eta$ in the term $X + U^Tb\eta^T q_r(X)$.  The latter expression suggests using the Ackermann formula for computation of $\eta$ in shifting the eigenvalues $M_{r}$ of $X$ to $L_r$:
\begin{equation}
\label{eq:poleplacement_gen_gen}
%\om_{r}^T =\eta^T P^T,~~\text{with}~~
\om_{n-r}^T =\eta^T U^T,\quad \eta^T= [0,\ldots,1]\,\mathcal{C}^{-1}(X,U^T b).
\end{equation}
In words, {if $\om_{n-r}$ is fixed perpendicularly to the invariant subspace $\msv$, then the corresponding open-loop eigenvalues remain unchanged} if we apply the feedback of the form \eqref{eq:poleplacement_gen_rer} with \eqref{eq:kappa} and \eqref{eq:poleplacement_gen_gen}. This fact provides a geometric interpretation for the term $\om_{n-r}$ in the expression \eqref{eq:poleplacement_gen_rer}.

With reference to \eqref{eq:geom1},  it is easily seen that the invertibiliy of the controllability matrix $\mathcal{C}^{-1}(X,U^T b)$ in \eqref{eq:poleplacement_gen_gen} requires
\[
\textit{rank}\, (U^T \, [b, Ab,\ldots, A^{r-1}b]) = r,
\]
which refers to the projected subsystem $(U^TAU,U^Tb)$ onto the subspace $\msu\subseteq\mathbb{R}^r$ \cite{Saad86}.

%
%
%%------------------------------------------------------------
\subsection{Sequential spectrum assignment}
\label{Sequential spectrum assignment}
Comment \emph{(iv)} indicates the difficulties with the invertibility
of the underlying controllability matrix, while in the previous
section we saw that the latter is reduced due to the projection of the
system matrix onto a subspace of a lower dimension. % in the context of partial spectrum assignment.
This idea can now be utilized sequentially as suggested by the
following algorithm. Let
\begin{align}
  \varLambda(A) = \bigcup_{\ell=1}^m M_{\ell}, \quad   \varLambda(\bar A) = \bigcup_{\ell=1}^m L_{\ell},
\label{eq:eigvals}
\end{align}
where $M_\ell$ includes a submultiset of self-conjugate open-loop
eigenvalues, and $L_\ell$ the corresponding desired self-conjugate
closed-loop eigenvalues. In other words, the eigenvalues in $M_\ell$
are to be shifted to $L_\ell$ for all $\ell\in\numm$. Then, introduce:
\begin{align}
  u_{\ell} = \omega^T_{\ell}q_{\ell}(\bar A_{\ell}) x +u_{\ell+1}, \quad \ell\in\numm
\label{eq:sequent}
\end{align}
with $u=u_{1}$, $u_{m+1}=0$, $\bar A_1 = A$, $\bar A_{\ell+l} = \bar A_{\ell}+ b \omega^T_{\ell}q_{\ell}(\bar A_{\ell})$,
\begin{align}
\label{eq:geom1}
{\bar A_\ell} = \begin{bmatrix} X_\ell & 0 \\ \ast & Y_\ell \end{bmatrix} : \begin{matrix} \mathcal{U}_\ell \\ \oplus \\ \mathcal{V}_\ell \end{matrix} \rightarrow \begin{matrix} \mathcal{U}_\ell \\ \oplus \\ \mathcal{V}_\ell \end{matrix},
\end{align}
where $\msv_\ell=\textit{Range}(V_\ell)$ represents the $\bar A_\ell$-invariant subspace corresponding to the eigenvalues $\varLambda(\bar A_\ell)\setminus M_\ell$, $\msu_\ell=\textit{Range}(U_\ell)$ is orthogonal to $\msv_\ell$ in $\mathbb{R}^n$,
\begin{align}
\om^T_{\ell} = \eta_\ell^T U^T_\ell,\quad \eta^T_\ell=  [0,\ldots,1]\,\mathcal{C}^{-1}(X_\ell,U_\ell^T b)
\label{eq:sequent}
\end{align}
and $q_\ell(\cdot)$ is the characteristic polynomial corresponding to the desired eigenvalues in $L_\ell$.
Effectively, we obtain: %the resulting feedback gain reads:
\begin{align}
 k^T = \sum_{\ell=1}^m \om^T_{\ell} q_\ell(\bar A_\ell).
\label{eq:sequent}
\end{align}
In words, the vector $\om_\ell$ is set perpendiculary to the invariant
subspaces $\msv_\ell$ corresponding to the unaltered eigenvalues at
the $\ell^\text{th}$ iteration, while the Ackermann formula is used to
design the feedback gain $\eta_\ell$ for the assignment of the eigenvalues
$\varLambda_\ell$ in the projected subspace. This procedure is repeated
sequentially. Thereby, $\bar A_\ell$ represents the closed-loop system
matrix up to the $\ell^\text{th}$ iteration. Finally, if
$M_\ell$ includes a pair of conjugated poles only, then this algorithm
reduces to the Ackermann's method of invariant planes \cite{ackermann93}.

\vspace{-5pt}
\section{Conclusion}
This short note introduces a slightly
generalized version of pole placement formulae and discusses its
relationships to Ackermann, Bass-Gura and Simon\,\&\,Mitter
algorithms. It extends and completes initial ideas of \cite{BajAJC}. The author thanks Dietrich Flockerzi for  useful discussions.%usability in a sequential pole placement algorithm.

\vspace{-5pt}

\bibliographystyle{plain}
\bibliography{bib/aigaion_fgrs_export_2012_05_30}

\end{document}